\documentstyle[12pt,epsf]{article}
\def\m{\multicolumn}
\date{}
\title{Optical Photometry of the GRB 010222 Afterglow}
\author{R. Cowsik$^1$, T.P. Prabhu$^1$, G.C. Anupama$^1$, 
B.C. Bhatt$^2$, D.K. Sahu$^3$, \\
S. Ambika$^1$, Padmakar$^3$, S.G. Bhargavi$^1$\\
$^1$Indian Institute of Astrophysics, II Block, Koramangala,\\
Bangalore 560 034, India\\
$^2$Indian Astronomical Observatory, IIA Leh Campus, \\ 
Indian Institute of Astrophysics,\\
Ladakh 194 101, India\\
$^3$Centre for Research and Education in Science and Technology, \\
Indian Institute of Astrophysics, Hosakote 562114, India}

\begin{document}
\maketitle
\begin{abstract}
The optical afterglow of GRB 010222 was observed using the recently
installed 2-m telescope at the Indian Astronomical Observatory, Hanle,
and the telescopes at the Vainu Bappu Observatory, Kavalur, beginning 
$\sim 0.6$~day after the detection of the event. The results based on 
these photometric observations combined with others reported in the 
literature are presented in this paper. The $R$ band light curve shows 
an initial decline of intensities proportional to $t^{-0.542}$ which 
steepens, after 10.3 hours, to $t^{-1.263}$.  Following the model of 
collimated outflow, the early break in the light curve implies a very 
narrow beam angle ($\sim 2^\circ-3^\circ$). The two decay rates are 
consistent with the standard jet model in a uniform density ambient 
medium, but require a hard spectrum of electron power density with 
$p \sim 1.5$. 
The $R$ band light between 14 and 17 hours since outburst
departs from the power law fit by 0.1 mag and shows some evidence for
fluctuations over timescales of an hour in the observer's frame. 
Such deviations are expected due to density inhomogeneities if the 
ambient medium is similar to the local interstellar medium. 
GRB 010222 is thus an example of a highly collimated outflow with a
hard spectrum of electron energy distribution in normal interstellar  
environment. 
\end{abstract}

\section{Introduction}

GRB 010222, one of the brightest bursts detected so far, was detected in
both the Gamma-Ray Burst (GRB) Monitor and the Wide Field Camera-1 (WFC)
aboard the BeppoSAX satellite, on Feb 22.308 (Piro 2001). The refined 
position from WFC was centred at 
${\rm{RA}} = 14^{\rm{h}} 52^{\rm{m}} 12.24^{\rm{s}}$ and 
${\rm{DEC}} = +43^\circ 00^\prime 43.2^{\prime\prime}$ (J2000) with an 
error circle of radius 2.5 arcmin (Gandolfi 2001a). BeppoSax detected 
the  X-ray afterglow in the 1.6--10~keV range about 9~hours after the 
GRB prompt event (Gandolfi 2001b). The X-ray afterglow was also observed
by the Chandra Observatory about 15 hours after the burst on Feb 22.936 
UT (Garmire et al. 2001) and again on March 3.54 UTC (Harrison, Yost \& 
Kulkarni 2001). The Bepposax observations between 8 and 65 hours since
the outburst are reported by in't Zand et al (2001).

Henden (2001a) reported the detection of the optical transient (OT) at
about 18th magnitude in $V$, about 4.4 hours after the burst, located at
$\rm{RA}=14^{\rm{h}} 52^{\rm{m}} 12.0^{\rm{s}}$, 
$\rm{DEC}=+43^\circ 01^\prime 06^{\prime\prime}$ (J2000). 
Photometry of the OT has been reported in several GCN circulars and by 
Masetti et al (2001), Lee et al (2001), Sagar et al (2001) and Stanek
et al (2001a).  The optical spectrum of the OT showed several absorption
line systems superimposed on a power-law continuum (Garnavich et al 
2001; Jha et al 2001a,b; Masetti et al 2001). These absorption line 
systems consisted of red-shifted doublets of Fe~II (2585, 2599 \AA), 
Mg~II (2796, 2803 \AA) and a host of other lines which indicate 
basically three intervening absorbers located at $z\sim$ 0.927, 1.155 
and 1.475. The detection of Mg~I line and the presence of strong 
absorption lines of Mg~II with equivalent widths of $\sim$ 1--3 \AA\ are
generally interpreted as indicating that the GRB occurred in a galaxy at
the highest redshift seen, viz. $z$ = 1.475. The equivalent width of the
Mg~II doublet is particularly large, especially in the system at $z$ = 
1.475 ($\sim$ 3 \AA) which implies a column density of $n_H\sim 10^{21}$
atoms cm$^{-2}$, if the composition is taken to be universal.  

We report here the optical observations made using the recently 
installed 2-m telescope at the Indian Astronomical Observatory, located 
in Hanle, Ladakh ($78^\circ 57^\prime 51^{\prime\prime}$~E;
$32^\circ 46^\prime 46^{\prime\prime} $~N; altitude of 4500~m
above msl). This set of observations forms one of the first data sets 
obtained from the telescope on scientific programmes even as the 
commissioning tests are underway. Also included in the paper are the 
observations made using the telescopes at the Vainu Bappu Observatory.

\section{Observations and Analysis}

\subsection{Observations}

The optical transient (OT) associated with GRB 010222 was observed for
$\sim 8$ days both from the Indian Astronomical Observatory (IAO) and 
the Vainu Bappu Observatory (VBO) beginning $\sim 0.6$ day after the
event. 
 
\subsubsection{IAO}
$R$ band CCD photometry from IAO was carried out using the optical CCD 
imager on the recently installed 2-m telescope. The telescope was 
installed in September 2000 and is undergoing commissioning tests. 
Details of the observatory and telescope may be obtained from 
\it{http://www.iiap.ernet.in/iao} \rm and also Anupama (2000). 
The CCD imager is equipped with Johnson-Cousins $UBVRI$ filters 
(see Bessell 1990) and a $1024\times 1024$ SITe chip of 24$\mu$m square
pixels covering a field of $4.7\times 4.7$~arcmin at the f/9 Cassegrain 
focus.  On Feb 22, four 600s exposures were obtained, while three 600s 
and two 300s exposures were obtained on Feb 26. The images recorded on 
Feb 22 are slightly trailed due to tracking and guiding errors, but are 
usable.

\subsubsection{VBO}

The OT was observed in the $R$ band on Feb 22 and 23 using the 1-m Zeiss
telescope equipped with a CCD imager similar to that in use at IAO. The
total field covered at the f/13 Cassegrain focus is $6.5\times 
6.5$~arcmin.

Using a CCD imager similar to the other two systems, at the f/3.25 prime
focus of the 2.3m VBT, the OT was observed in the $R$ and $I$ bands on 
Feb 24 and 25 and in the $R$ band on Feb 28 and Mar 1 (Cowsik \& 
Bhargavi 2001). The total field covered was $10.4\times 10.4$~arcmin.

\subsection{Analysis}

We describe in the following, the analysis of the IAO and VBO 1-m data.
The details of the VBT data analysis are presented in Bhargavi (2001).

All images were bias subtracted using the mean value of the nearest bias
frames and flatfield corrected using a master flat generated by 
combining several twilight flats. Often a few processed frames were 
aligned and combined before extracting magnitudes in order to improve 
the S/N ratio.  These frames are indicated in a single row of Table 1.
It was not possible to perform a PSF fitting to the data obtained from 
IAO on Feb 22 as the stellar profiles are affected by the tracking 
and guiding errors.  In order to have a uniform mode of magnitude
estimation for all sets of data, magnitudes were obtained using aperture
photometry and growth curve method (Howell 1989). Aperture photometry of
the OT and several bright stars in the field was performed using 
concentric apertures ranging from 0.7 to 14 arcsec diameter centred on 
the object.  Sky background was subtracted using a 1.5 arcsec annulus 
with an inner radius of 5--8 arcsec for the OT and 8--10 arcsec for the 
brighter stars.  Based on a growth curve for the brighter stars, a 
6--7 arcsec aperture was chosen for the photometric measurements to 
include most of the light from the source. The OT being faint, its 
magnitude estimated from the 3--4 arcsec aperture. Corrections for the 
photons from the source outside this aperture were made using a 
correction factor estimated from the growth curve of the brighter stars 
(average of Henden 5, 16, 18, 27).  The zero points were determined 
using star `A' of Stanek et al. (2001b) whose magnitude was assumed to 
be $R=17.175$ following Henden (2001b). 

The magnitudes of other stars in the field were compared with Henden's 
photometry to ascertain that no systematic errors exist between the two 
data sets.  A colour transformation equation was determined for 
residuals of the field stars with respects to Henden's photometry using 
Henden 5, 9, 11, 13, 16, 25, 33 (VBO) or 7, 11, 13, 16, 18, 20 (IAO).
The bluest star Henden 18 was not used at VBO since it showed that the
colour equation is nonlinear below $(V-R)=0.3$. Since the OT had a
colour of $(V-R)=0.37\pm 0.05$ (Masetti et al 2001), the nonlinearity 
does not affect the results. The zero-point correction to the field star
photometry, together with the colour correction, resulted in a total 
correction ranging between 0.014 and 0.022 mag for different frames.

The details of observations and derived magnitudes are listed in Table 1
and plotted in Fig.\ 1.

\subsection{$R$ band light curve}

The $R$ band light curve of the OT of GRB 010222 is obtained using 
magnitudes from the following sources: (1) IAO and VBO magnitudes 
reported in this paper (VBO1 = 1-m telescope; VBT = 2.34-m telescope), 
(2) Sampurnanand Telescope (ST) observations reported by Sagar et al. 
(2001), (3) Telescopio Nazionale Galileo (TNG), Asiago, and Loraino 
Observatory measurements reported by Masetti et al. (2001), (4) F.L. 
Whipple Observatory (FLWO) and Vatican Advanced Technology Telescope 
(VATT) measurements reported by Stanek et al. (2001a and kindly made 
available by K.Z. Stanek), and (5) Canada-France-Hawaii Telescope (CFHT)
measurements reported by Veillet (2001a, b), (6) Subaru telescope 
measurements reported by Watanabe et al (2001) --- though these 
estimates have large errors, they provide a point in the gap near the 
break. All these magnitudes are listed in Table 2.

\section{Results and Discussion}

\subsection{Power-law Fits}

The $R$ light curve shows a broken power-law behaviour with the first 
phase ending around 10 hours after which the decline became steeper. 
The curves before and after the break can be fit by power laws  
$$ I(R,t) = I_0(t/t_b)^{-\alpha}, $$ 
\noindent where the time $t$ is measured from the time of GRB detection,
{\it i.e.\ }2001 February 22.3080 (Piro 2001), and $t_b$ is the time of 
break in the curve. The power law exponents for the two phases are 
denoted by $\alpha_1\ (t<t_b)$ and $\alpha_2\ (t>t_b)$.

Two linear $\chi^2$-fits were obtained against $\log t$ using the 
magnitudes in Table 2. In order to obtain the asymptotic behaviour, the
points close to the break between $t$ = 0.3 -- 0.8~d were excluded. The
derived parameters are listed in Table 3 as Model 1.  The quoted errors 
for $I_0$ and $t_b$ are derived using $\pm 1\sigma$ errors on $\alpha_1,
\alpha_2$, and the intercepts in magnitude. The degrees of freedom 
($\nu$) are also shown. The fits are shown in Fig.~2 by straight lines.

The following smooth empirical function (cf. Beuermann et al. 1999) is
often used to fit light curves of optical transients of GRBs.

$$I(t)=(2^{\frac{1}{s}}I_0)\big[(t/t_b)^{\alpha_1s}+(t/t_b)^{\alpha_2s}\big]^{-\frac{1}{s}}$$

\noindent where $s>1$ is a sharpness parameter and increases as the
break becomes sharper.  We find that such a curve does not describe the 
data in Table 2 well.  We obtain a good fit if we exclude the region 
$t$ = 0.3 -- 0.8 d as shown in Model 2 of Table 3. The $\chi^2$ reduces
considerably as $s$ increases until $s\sim 10$ and decreases very slowly
thereafter. We have shown the fit for $s=10$ by a 
dotted curve in Fig.~2.  The constraint on smoothness also tends to 
increase the slope before the break and decrease it after the break.
While the smooth fit provides smaller formal errors on the parameters, 
there is no specific reason to prefer the asymptotic values resulting 
from this fit to the values derived by linear fits away from the break. 
Figure 2 shows systematic deviation of the points away from the break 
which can be reduced only by increasing the value of $s$. On the other 
hand, a large value of $s$ recovers the slopes fit by separate straight 
lines.  

The light curve parameters derived above may be compared with the 
values derived by Masetti et al. (2001): $\alpha_1 = 0.60\pm 0.03$,
$\alpha_2=1.31\pm 0.03$, $t_b = 0.48\pm 0.02$ d, and $s=3.4\pm 0.8$. 
Sagar et al (2001) derive a slightly steeper fit with $\alpha_1 = 0.76
\pm 0.03$, $\alpha_2 = 1.37\pm 0.02$ with $s=4$, and the break $t_b = 
0.71\pm 0.07$ considerably later than derived here, but in agreement 
with Stanek et al (2001a): $\alpha_1 = 0.80\pm 0.05$, $\alpha_2 = 1.30
\pm 0.05$, $t_b = 0.72\pm 0.1$, though with a much sharper break 
($s=10$). 

There appears a drop in the light curve with respect to both the fits
around Feb 23.0.  The observations of IAO and UPSO both agree in this 
behaviour (see inset in Fig.~1). The VBO 1-m magnitudes have large 
errors and the deviations are not apparent.

\subsection{The Jet Model and Physical Conditions}

The favoured model for GRB afterglows is synchrotron or inverse
Compton scattering of electrons accelerated in a relativistic shock
wave expanding in its surrounding medium (M\'esz\'aros \& Rees 1997;
Sari, Piran \& Narayan 1998). The shock decelerates as it sweeps up 
ambient matter and the emission fades down.  A steepening of the light 
curve has been observed in the optical transients of several GRB sources
and has been generally explained either through expansion in a dense 
medium or in terms of collimated outflows (Rhoads 1999; Sari, Piran \& 
Halpern 1999).  The latter (jet) model is of special interest since it 
reduces the total energy requirement and increases the actual rate of 
occurrence of such events. According to this model, the hydrodynamics of
the relativistic jet is not influenced by the finite angular size during
the early evolution when the bulk Lorentz factor $\gamma >\theta_0^{-1}$
where $\theta_0$ is the opening angle of the jet in radians. In this 
case, the light curve behaves similar to the isotropic expansion. 
However, when $\gamma \sim \theta_0^{-1}$, the sideways expansion of the
jet becomes significant and the light curve becomes steeper. The break 
in the light curve is expected at 
$$t_j \sim 6.2(E_{52}/n_1)^{\frac{1}{3}}(\theta_0/0.1)^{\frac{8}{3}}\ \rm{hr},$$
\noindent
where $E_{52}$ is the isotropic energy of the ejecta in the units of
$10^{52}$ ergs, $n_1$ is the ambient particle density in cm$^{-3}$,
assumed to be homogeneous (Sari, Piran \& Halpern 1999). With 
$t_b=(1+z)t_j$ in the observer's frame, we derive
$$\theta_0 = 5^\circ .7\bigg[\frac{t_b}{6.2(1+z)}\bigg]^{\frac{3}{8}}
\bigg(\frac{n_1}{E_{52}}\bigg)^{\frac{1}{8}}.$$ 
For an isotropic energy of $E_{52}=78$ (in't Zand et al 2001), the 
observed value of $t_b =10.3\pm 2.6$ h, implies a beam angle in the 
range of $2^\circ .8\pm 0^\circ .3$ ($n_1=1$) and $16^\circ \pm 1^\circ 
.5$ ($n_1=10^6$). Note that the dependence on $n_1$ and $E_{52}$ is 
rather weak. 

The narrow beam angle ($3^\circ$ for $n_1=1$) implies that such 
outbursts actually occur at a rate 1500 times higher than observed.  
This is not much higher than inferred  for average GRBs with slightly 
wider beam angles, which has already triggerred an interest to look for 
their remnants in the local universe (Paczy\'nski 2001).

The models of the evolution of the outflow also predict a spectral
energy distribution (SED) with a peak either at the characteristic
synchrotron frequency ($\nu_m$) or at the cooling frequency ($\nu_c$) 
depending on whether the cooling is slow or fast (Sari \& Esin 2001). 
The spectrum would be self-absorbed below a frequency $\nu_a$ which 
generally lies in the radio region. The shape of the spectrum and its 
time evolution can be determined by a single parameter $p$, the electron
power law distribution index, assuming adiabatic expansion, or expansion
of a jet in a homogeneous medium and 
neglecting the dust extinction.  The shape and decay rate of the optical
spectrum would then depend on its location with respect to $\nu_c$ and 
$\nu_m$. We assume in the following that the spectrum is dominated by 
synchrotron rather than inverse Compton scattering and reached 
the slow cooling phase well before the break in the light curve.

Kulkarni et al (2001) report a 350 GHz detection of the OT on Feb 22.54 
at $4.2\pm 1.2$ mJy. It was detected at 22 GHz a little later (Feb 
22.62) with a flux of $0.70\pm 0.15$ mJy (Berger \& Frail 2001). The $R$
band flux at these times based on our fit are: 0.110 mJy (Feb 22.54) and
0.093 mJy (Feb 22.62). The 220 GHz detections on Feb 23 and 24 (Kulkarni 
et al 2001) are 
slightly lower than expected from these two points.  Based on these 
observations, $\nu_m$ is estimated to be in the sub-mm region just 
before the break. Further, the 350 GHz as well as 220 GHz flux remained 
nearly constant between Feb 22.54 and 24.68 (Kulkarni et al 2001). Such 
a behaviour is expected when the peak frequency is passing through this 
spectral region.  Kulkarni et al (2001) hypothesized that most of the
sub-mm flux arises from the host galaxy, either from starburst regions, 
or due to reprocessing of GRB radiation. In such an event, the sub-mm 
flux of the OT will be lower and the peak could be placed at higher
frequencies. 

It has been suggested by Lee et al (2001) that the break in $R$ band
decay may be caused by the passage of $\nu_m$ through this spectral
region.  The $JK$ magnitudes reported by Masetti et al (2001) around
1--2 days after the outburst show little variation with time.  If we 
suppose that $\nu_m$ was passing through this region around day 1.0, 
considering that it drops as $t^{-3/2}$ in the adiabatic expansion 
phase, it could have passed through the optical region around the
time of the break. However,  the light curves of Masetti et al (2001) 
appear fairly achromatic despite limited data in the early phase and 
thus there is no strong indication of $\nu_m$ passing through the 
optical region.   

If we consider the possibility that $\nu_m<\nu_{opt}<\nu_c$, we expect 
$\alpha_1 = 3(p-1)/4$ and $\alpha_2 = p$ (Sari, Piran \& Halpern 1999). 
The observed slopes yield two estimates of $p= 1.72\pm 0.09$ and $1.26
\pm 0.01$, considerably harder than the value of 2.4 obtained for 
several other GRBs. The optical spectrum is then predicted to vary as 
$\nu^{-\beta}$ with $\beta=(p-1)/2$. The two estimates of $\beta$ would 
then be $0.36\pm 0.05$ and $0.131\pm 0.005$, considerably flatter than 
observed.  

If we assume, on the other hand, that $\nu_m<\nu_c<\nu_{opt}$, we obtain
$\alpha_1 = (3p-2)/4$ and $\alpha_2=p$. The derived values of $p$ are 
$1.39\pm 0.09$ and $1.26\pm 0.01$ with the spectral index $\beta = p/2 =
0.70\pm 0.05$ and $0.631\pm 0.005$. These values are more consistent 
with each other and can be reconciled with the observed optical spectrum
considering that extinction in the host galaxy would modify this 
spectral region (Lee et al 2001). 

The 350 GHz flux decayed from 
$4.2\pm 1.3$ mJy on Feb 24.68 (Kulkarni et al 2001) to $0.7\pm 0.15$ mJy
on March 2 (Ivison et al 2001). If all the flux is attributed to the GRB,
the deduced decay index is $\alpha$ = 
1.52 which is numerically equal to $p$ in the jet model. The 2--10 keV 
flux also decayed with $\alpha$ = 1.33 beginning a couple of hours 
before the break to 2.4 days after the break (in't Zand et al 2001).

The expressions for the decay rate exponents derived by Sari, Piran \&
Halpern (1999) assume an electron power law index of $\sim 2.4$. 
Bhattacharya (2001) has derived expressions for the case of $p=$
1--2 assuming that there is an upper cutoff in the electron energy
spectrum at $\gamma_u \propto \Gamma^q$, where $\Gamma$ is the bulk
Lorentz factor of the blast wave, and a single power law between
$\gamma_m$ and $\gamma_u$. The decay rate exponents can then be derived 
in terms of $p$ and $q$. Assuming that $\nu_c <\nu_{\rm {opt}} < \nu_u$ 
where $\nu_u$ is another critical frequency associated with $\gamma_u$, 
Eqs (26) and (28) of Bhattacharya (2001) imply $3\alpha_2 = 4\alpha_1 
+2.$ Since $\alpha_2$ has been determined well, we derive $\alpha_1=
0.447$ using this relation, and derive a value of $q\sim 2$ for 
$p\sim 1.5$.

Dai \& Cheng (2001), on the other hand, assumed a flat electron energy 
spectrum that gets steeper after the cooling break at $\gamma_c > 
\gamma_m$, and deduced that the light curve of the jet would decay
as $t^{-2}$ irrespective of the value of $p$ as long as it is between
1--2. Observations are at variance with this model suggesting that 
either the jet model is not applicable, or the electron energy spectrum 
has a different shape.

It thus appears likely that the electron energy spectrum was 
considerably harder ($p\le 1.5$) for GRB 010222 compared to typical GRBs
($p\sim 2.0-2.5$) exhibiting steeper decay in the optical. Sagar et al 
(2001) also reached a similar conclusion from the spectral slope 
assuming $\nu_c < \nu_{\rm{opt}}$.  A similar hard spectrum was 
suggested by Panaitescu (2001) for GRB 000301c whose spectrum around the
light curve break shows a good resemblance with GRB 010222.


\subsection{The Light Curve soon after the Break}

The light curve of GRB 010222 is not well-observed between the time
it was lost to the Pacific coast and later when it was recaptured from
the Indian observatories. The early observations from IAO and ST suggest
that the light curve had a significant dip with respect to the power
law after the break. Similar deviations were also seen in the case of 
GRB 000301c and were attributed by Berger et al (2000) to the 
inhomogeneities in the ambient interstellar medium. These authors have 
argued that the flux above $\nu_a$ would vary with ambient density as 
$n_1^{1/2}$.  Thus the brightness would increase when the shock
overtakes denser regions and dip when it reaches lower density regions.

Wang \& Loeb (2000) and Ruffini et al. (2001) have considered the 
effects of inhomogeneities in the interstellar matter on the radiation 
from the afterglow in somewhat greater detail. They find that variations
are expected over timescales of seconds or tens of seconds when the bulk 
Lorentz factor is 
high ($> 100$) and increase to days when the Lorentz factor reduces to 
$\sim 3$, if one assumes that the shock propagates through a medium 
similar to the local interstellar matter. 

The observed $R$ magnitudes of GRB 010222 between 14 and 17 hours since 
the outburst depart from the power law by $0.107\pm 0.066$ mag including
all the data in Table 2, and $0.111\pm 0.055$ mag if we consider only 
the IAO and ST estimates which have an average error of 0.027 mag (see 
inset in Fig.~1). The 0.1 mag depression in the light curve can easily 
be caused by a 20\% decrease in the ambient density. The rms fluctuation
of 0.06 mag, though statistically not very significant, can be caused by 
density fluctuations of little over 10\%.  The IAO and ST observations 
have a mean interval of 0.6 hours.  The bulk Lorentz factor has a value 
of about 20 at this epoch, at which time, density fluctuations of 
$\sigma_n/n$ of 1--3 over 100 AU length scales and 0.1--0.2 over 1000 AU
length scales can produce fluctuations of 0.05 mag amplitude over 
timescales of an hour.  The values of density contrast are consistent 
with the inhomogeneities in the local interstellar medium (Lauroesch 
\& Meyer 1999).  The observed column densities of $N_{\rm{H}}\sim 
10^{21}$ cm$^{-2}$ imply a total path length of 300~pc for $n_1\sim 1$, 
comparable to the thickness of disks of galaxies. We thus suspect that
GRB 010222 occurred in a normal interstellar environment.

Fluctuations are not noticed in the light curves of most GRB afterglows 
suggesting that the ambient medium is considerably smooth around them. 
On the other hand, structure is evident over timescales of tens of 
seconds in the BATSE observations (see in't Zand et al in case of GRB 
010222) which can be attributed to interstellar density fluctuations 
(Ruffini et al. 2001).  Future GRB afterglows may be monitored more
systematically with a view to probing the stucture of the ambient 
medium. 

\section{Summary}

The $R$ band photometry of the OT of GRB 010222 performed with the 
telescopes at IAO, Hanle and VBO, Kavalur is presented here. These 
measurements together with the photometry reported in the literature 
show that the $R$ band light curve of GRB 010222 had a sharp break at
10 hours from the outburst with decay rates 0.542 and 1.263 before and
after the break. The light curve shows some evidence of fluctuations  
soon after the break. It is suggested that the variations are caused by 
the inhomogeneities in the ambient medium.

Following the standard models of collimated jet outflow, the early break
in the light curve implies that the beam had a very narrow opening angle
of $\sim 3^\circ$. This is probably the narrowest observed for any GRB 
so far.

The decay rates imply a hard spectrum of electron energy distribution 
($p\sim 1.5$), which is consistent with the a spectral index of $\beta 
\sim 0.75$ in the optical region.

It would thus appear that GRB 010222 is an example of a high-energy
outburst with a narrow jet with a hard electron energy spectrum in a 
normal interstellar medium. 

\subsection*{Acknowledgements}

We are greatly indebted to the 2-m telescope project team which 
helped in the procurement and installation of the telescope at the
high altitude site in less than 3 years, and developed all the
necessary infrastructure including the enclosure, power supply,
liquid nitrogen supply, and communication link to Hosakote
near Bangalore for remote operation and downloading the data.
We also thank K. Jayakumar for obtaining images from VBT on February 28
and March 1. We thank P. Bhattacharjee and P.N. Bhat for stimulating
discussions and the referees for their valuable comments.

\newpage
\begin{center}
\epsfxsize=16cm
\leavevmode
\epsfbox{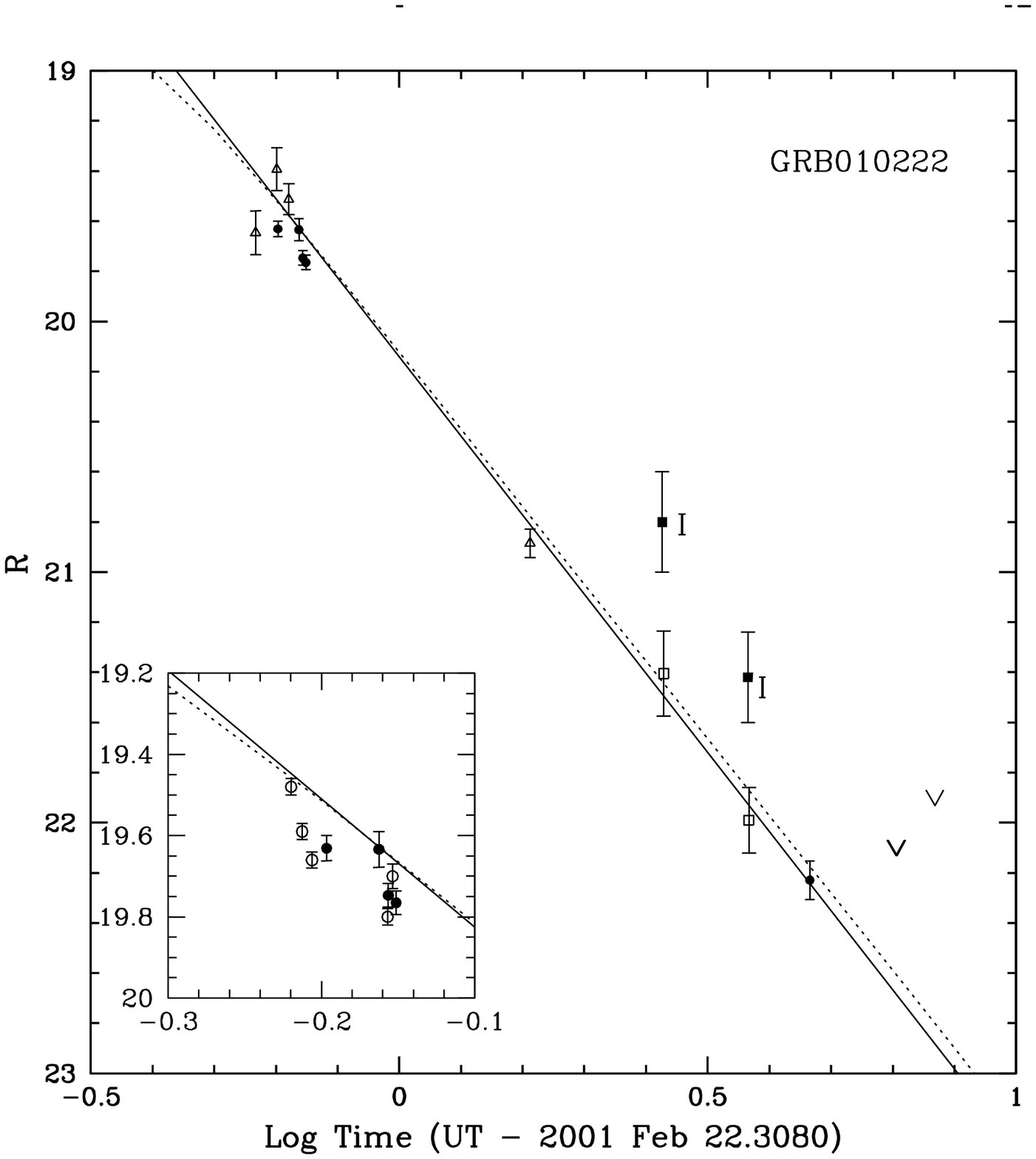}
\end{center}
\vskip 5mm\noindent
{\bf Figure 1.} Photometric observations reported in this paper. 
Magnitudes are in $R$ band except where noted.  Circles: IAO 2-m 
telescope; triangles: VBO 1-m telescope; open squares: VBT ($R$ band); 
filled squares: VBT ($I$ band). The $R$ band upper limits obtained 
with VBT are also shown.  The fits shown in Fig.~2 are also marked.
The inset shows the region soon after the break magnified, with the 
IAO (filled circles) and ST (Sagar et al 2001; open circles) magnitudes.

\newpage
\begin{center}
\epsfxsize=16cm
\leavevmode
\epsfbox{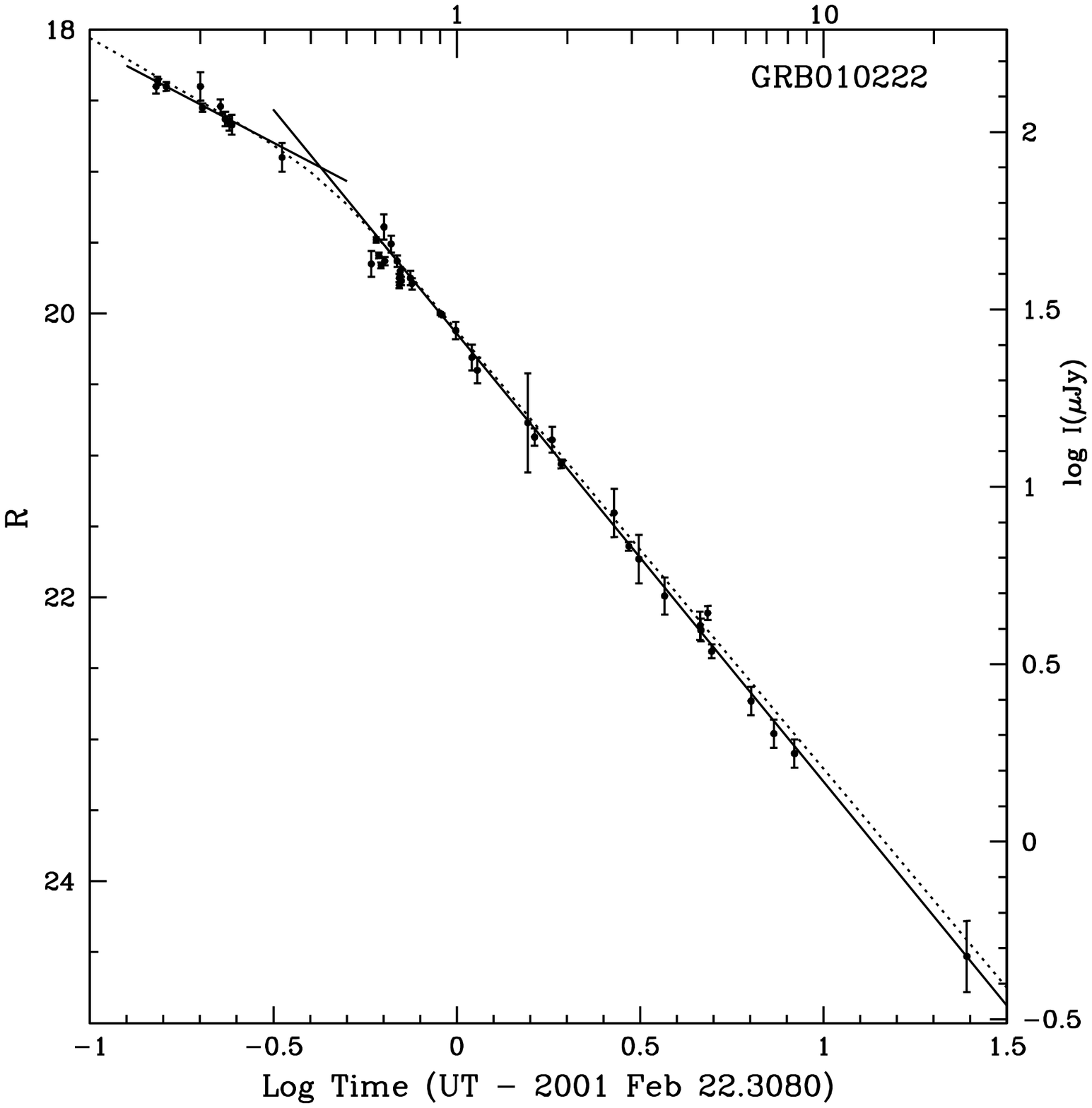}
\end{center}
\vskip 5mm\noindent
{\bf Figure 2.} The broken power law fits to the early and late phases
of the GRB 010222 optical transient in the $R$ band corresponding to
Model 1 of Table 3. A smooth fit corresponding to Model 2 of Table 3
is shown as a dotted curve. 

\newpage
\begin{tabular}{rlccl}
\m{5}{l}{{\bf Table 1.} GRB010222: CCD Photometry}\\
\hline\\
Date (UT)  & Telescope & Exp Time(s) & Band & Magnitude\\
\hline\\
2001 Feb \\
22.8931 & VBO 1-m & 180+300 & $R$ & $19.646 \pm 0.087$\\
22.9413 & VBO 1-m & 1800 & $R$ & $19.393  \pm 0.086$\\
22.9438 & IAO 2-m & 600 & $R$ & $19.631 \pm  0.031$ \\
22.9703 & VBO 1-m & 1800 & $R$ & $19.512  \pm 0.062$\\
22.9958 & IAO 2-m & 600 & $R$ & $19.634 \pm  0.044$ \\
23.0056 & IAO 2-m & 600 & $R$ & $19.747 \pm  0.029$\\
23.0139 & IAO 2-m & 600 & $R$ & $19.765 \pm  0.029$\\
23.9375 & VBO 1-m & 2400+1800+1200 & $R$ & $20.870 \pm  0.057$\\
24.977 & VBT & 600 & $I$ & $20.8 \pm  0.2$\\
24.9924 & VBT & 600 & $R$ & $21.405 \pm 0.17$\\
25.9816 & VBT & 900 & $I$ & $21.42 \pm  0.18$\\
25.9955 & VBT & 900 & $R$ & $21.99 \pm  0.13$\\
26.9385 & IAO 2-m & 3$\times$600+2$\times$300 & $R$ & $22.229 \pm  0.077$\\
28.9014 & VBT & 1200 & $R$ & $>22.1$\\
28.9257 & VBT & 2400 & $R$ & $>22.1$\\
2001 Mar \\
1.9319 & VBT & 1800 & $R$ & $>21.9$\\
\hline\\
\end{tabular}

\newpage
\begin{tabular}{rll}
\m{3}{l}{{\bf Table 2.} GRB010222: $R$ magnitudes used}\\
\hline\\
Date (UT)  & Telescope & Mag \\
\hline\\
2001 Feb\\
22.4595 & FLWO& $18.40\pm  0.05$\\      
22.4610 & FLWO& $18.36\pm  0.03$\\      
22.4695 & FLWO& $18.40\pm  0.03$\\     
22.508  & SUBARU&$18.4\ \pm 0.1$\\
22.5106 & FLWO& $18.55\pm  0.03$\\    
22.5349 & FLWO& $18.54\pm  0.05$\\   
22.5417 & FLWO& $18.63\pm  0.05$\\  
22.5480 & FLWO& $18.66\pm  0.05$\\ 
22.5519 & FLWO& $18.67\pm  0.07$\\
22.642 & SUBARU&$18.9\ \pm 0.1$\\
22.8931 & VBO1& $19.65\pm  0.09$\\ 
22.911  & ST  & $19.48\pm  0.02$\\
22.921  & ST  & $19.59\pm  0.02$\\   
22.930  & ST  & $19.66\pm  0.02$\\       
22.9413 & VBO1& $19.39\pm  0.09$\\
22.9438 & IAO & $19.63\pm  0.03$\\
22.9703 & VBO1& $19.51\pm  0.06$\\
22.9958 & IAO & $19.63\pm  0.04$\\
23.005  & ST  & $19.80\pm  0.02$\\       
23.0056 & IAO & $19.75\pm  0.03$\\
23.010  & ST  & $19.70\pm  0.03$\\       
23.0139 & IAO & $19.77\pm  0.03$\\
23.056  &Masetti&$19.75\pm  0.05$\\
23.063  &Masetti&$19.79\pm 0.04$\\
23.211	&Masetti&$20.00\pm 0.01$\\	
23.219	&Masetti&$20.01\pm 0.01$\\    
23.3015	& FLWO& $20.12\pm  0.06$\\
23.4076	& FLWO& $20.31\pm  0.09$\\
23.4458	& FLWO& $20.40\pm  0.09$\\
23.870  & ST  & $20.77\pm  0.35$\\    
23.9375 & VBO1& $20.87\pm  0.06$\\
24.127 &Masetti&$20.89\pm  0.09$\\
24.236 &Masetti&$21.06\pm  0.03$\\
24.239 &Masetti&$21.05\pm  0.02$\\
24.9924 & VBT & $21.41\pm  0.17$\\ 
\hline\\
\end{tabular}

\newpage

\begin{tabular}{rll}
\m{3}{l}{{\bf Table 2.} contd.}\\
\hline\\
Date (UT)  & Telescope & Mag \\
\hline\\
2001 Feb\\
25.253 &Masetti&$21.64\pm  0.03$\\
25.4469	& FLWO& $21.73\pm  0.17$\\	
25.9955 & VBT & $21.99\pm 0.13$\\ 
26.911  & ST  & $22.20\pm  0.10$\\
26.9385 & IAO & $22.23\pm  0.08$\\
27.139 &Masetti&$22.11\pm 0.25$\\
27.264 &Masetti&$22.38\pm  0.05$\\
28.653  & CFHT& $22.73\pm 0.10$\\	
2001 Mar \\
1.628  & CFHT& $22.96\pm 0.10$\\	
2.641	& CFHT& $23.10\pm 0.10$\\	
18.9	& VATT& $24.53\pm 0.25$\\	
\hline\\
\end{tabular}

\vskip 15mm


\begin{tabular}{llll}
\m{4}{l}{{\bf Table 3.} Light curve fit parameters}\\
\hline\\
Parameters & Model 1 & Model 2 & Model 3\\
\hline\\
$I_0$      & $79^{+35}_{-21}$ & $77.1\pm0.4$&$77.6\pm0.4$\\
$t_b (d)$  &$0.427\pm0.107$   & $0.434\pm0.003$&$0.431\pm0.002$\\
$\alpha_1$ &$0.542\pm0.071$   & $0.594\pm0.011$&$0.598\pm0.008$\\
$\alpha_2$ &$1.263\pm0.011$   & $1.234\pm 0.005$&$1.241\pm0.003$\\
$s$        & --- & 10 & 4.5\\
$\chi^2(\nu)$ & 4.5(7), 21(20) &39 (27) & 126 (42)\\
\hline\\
\end{tabular}
\end{document}